\documentclass[final]{elsarticle}

\date{September 12, 2016}
\usepackage{lineno,hyperref}
\modulolinenumbers[1]

\usepackage{graphicx}

\journal{Journal of Volcanology and Geothermal Research}

\biboptions{authoryear}

\begin{document}

\begin{frontmatter}

\title{Computer Aided Detection of Transient Inflation Events at Alaskan Volcanoes using GPS Measurements from 2005-2015}

\author[haystack]{Justin D. Li\corref{mycorrespondingauthor}}
\cortext[mycorrespondingauthor]{Corresponding author}
\ead{jdli@haystack.mit.edu}

\author[haystack]{Cody M. Rude}
\author[haystack]{David M. Blair}
\author[haystack]{Michael G. Gowanlock}
\author[mit]{Thomas A. Herring}
\author[haystack]{Victor Pankratius}

\address[haystack]{Astro-\&Geo-Informatics Group, MIT Haystack Observatory, Westford, MA USA}
\address[mit]{Department of Earth, Atmospheric and Planetary Sciences, MIT, Cambridge, MA, USA}

\begin{abstract}
Analysis of transient deformation events in time series data observed via networks of continuous Global Positioning System (GPS) ground stations provide insight into the magmatic and tectonic processes that drive volcanic activity. Typical analyses of spatial positions originating from each station require careful tuning of algorithmic parameters and selection of time and spatial regions of interest to observe possible transient events. This iterative, manual process is tedious when attempting to make new discoveries and does not easily scale with the number of stations. Addressing this challenge, we introduce a novel approach based on a computer-aided discovery system that facilitates the discovery of such potential transient events. The advantages of this approach are demonstrated by actual detections of transient deformation events at volcanoes selected from the Alaska Volcano Observatory database using data recorded by GPS stations from the Plate Boundary Observatory network. Our technique successfully reproduces the analysis of a transient signal detected in the first half of 2008 at Akutan volcano and is also directly applicable to 3 additional volcanoes in Alaska, with the new detection of 2 previously unnoticed inflation events: in early 2011 at Westdahl and in early 2013 at Shishaldin. This study also discusses the benefits of our computer-aided discovery approach for volcanology in general. Advantages include the rapid analysis on multi-scale resolutions of transient deformation events at a large number of sites of interest and the capability to enhance reusability and reproducibility in volcano studies.
\end{abstract}

\begin{keyword}
Alaska; Volcanoes; Transient Inflation Events; GPS; Computer-Aided Discovery
\end{keyword}

\end{frontmatter}

\section{Introduction}

Volcanology has greatly benefited from continued advances in sensor networks that are deployed on volcanoes for continuous monitoring of various activities and events over extended periods of time. Currently the scientific community is expanding and enhancing such sensor networks to improve their spatial and temporal resolutions, thus creating new opportunities for conducting large scale studies and detecting new events and behaviors \citep{Dzurisin2006volcano, Ji2011}.

However, the increasing amount of data recorded by geodetic instruments poses formidable challenges in terms of storage, data access, and processing. This trend makes manual detection and analysis of new discoveries increasingly difficult. Large data sets also drive the need for cloud data storage and high-performance computing as local machines gradually become less capable of handling these computational workloads. Therefore, more sophisticated computational techniques and tools are required to address to this challenge.

To expand current toolsets and capabilities in volcanology, this article presents a computer-aided discovery approach to volcanic time series analysis and event detection that helps researchers analyze the ever-increasing volume and number of volcanic data sets. A key novel contribution of this approach is incorporating models of volcanology physics to generate relevant and meaningful results. We demonstrate its applicability by examining geodetic data from the continuous Global Positioning System (GPS) network operated by the Plate Boundary Observatory (PBO; http://pbo.unavco.org), and through the detection and analysis of transient events in Alaska. 

Our computer-aided discovery system implements a processing pipeline with configurable, user-definable stages. We can then select from a number of algorithmic choices and allowable ranges for numerical parameters that define different approaches to volcanic data processing. Using these specifications, the system employs systematic techniques to generate variants of this pipeline to identify transient volcanic events. Variations in the processing pipeline -- the choice of filters, the range of parameters, and so forth -- can highlight or suppress features of a transient event and alter the certainty of its detection. As the optimum pipeline variant is not usually known a priori and is typically selected manually, a tool that automates this search helps make the discovery process more efficient and scalable. Additionally, once configured, these software pipelines can be reused for other data sets -- either to test and reproduce results from past studies, or to detect new or previously unknown events.

The article is organized as follows. Section 2 introduces the different geodetic data sets used in this study and briefly outlines the geographic region of interest. Section 3 describes the idea of a computer-aided framework and the specific implementation developed and utilized for this study. Section 4 presents the detection and validation of three transient deformation events, whose physical characteristics are discussed in Section 5, along with a discussion of the requirements and advantages of the discovery pipeline. Finally, Section 6 concludes with the contributions of the developed computer aided system and suggestions for further application and refinements.

\section{Geodetic Data}

Continuous GPS position measurements (``time series'' data) have become a particularly useful source of geodetic data in the past several decades with millimeter-scale resolution and wide-scale deployment. Here, we analyze daily GPS time series data collected by averaging 24 hours of recorded measurements at PBO stations in Alaska between 2005 and 2015. The GPS data are obtained from the PBO archives as a Level 2 product describing GPS station time series positions. This data is generated from Geodesy Advancing Geosciences and EarthScope GPS Analysis Centers at Central Washington University and New Mexico Institute of Mining and Technology and combined into a final derived product by the Analysis Center Coordinator at Massachusetts Institute of Technology. Locations of Alaskan volcanoes along with logs of monitored volcanic activity are obtained from the \citet{AVO_web} (AVO; http://avo.alaska.edu) and used to provide regions of interest that bound subsets of the PBO GPS stations.

\begin{figure}[h]
  \centering
    \includegraphics[width=\textwidth]{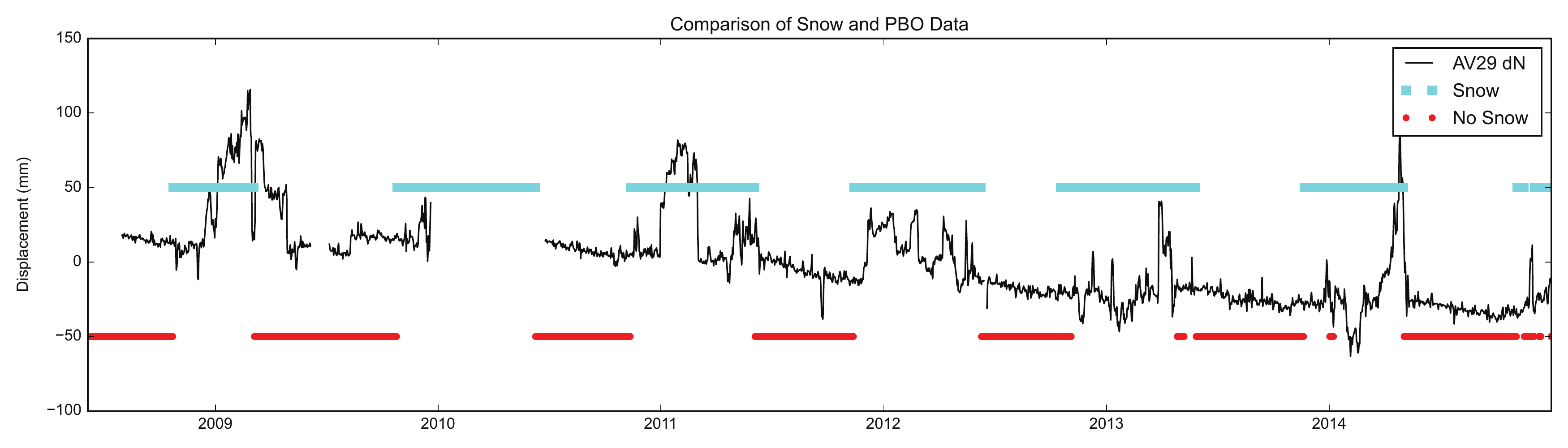}
    \caption{A comparison of the NSIDC snow cover data overlaid on the PBO data showing how the NSIDC data can be used to identify regions of high variability and missing PBO data. The blue points (upper line) mark days with snow and the red points (lower line) mark days without snow at station AV29, with its northward motion plotted for comparison in black.}
    \label{fig:Fig1}
\end{figure}

In addition to the GPS data, we also utilize snow cover data from the National Snow and Ice Data Center (NSIDC; http://nsidc.org) to address possible data corruption due to accumulated snow on GPS antennas, which can be significant as shown in Figure~\ref{fig:Fig1}. This novel data fusion of GPS time series data with the NSIDC National Ice Center's Interactive Multisensor Snow and Ice Mapping System Daily Northern Hemisphere Snow and Ice Analysis 4 km Resolution data product \citep{NSIDC_ref} helps identify stations with corrupted data resulting from seasonally occurring errors. To reduce the computational time and complexity, we apply a preprocessing step that combines NSIDC snow mapping data with the PBO GPS data as a single data product, which we use in later analyses. This data fusion approach takes a first step towards solving the challenging problem of resolving data corrupted by complex snow conditions.

\section{The Computer-Aided Discovery Pipeline}

We demonstrate the utility of our approach for volcanology by analyzing GPS measurements of displacements near and at volcanoes in Alaska between 2005 and 2015. We evaluate 137 volcanoes to determine sites with a sufficient number of GPS stations and time coverage and then analyze the selected volcanoes for potential transient signals indicative of inflation events. We find 3 transient signals, consistent with volcanic inflation, between 1/1/2005 and 1/1/2015 at Akutan, Westdahl, and Shishaldin volcanoes in the Alaskan Aleutian Islands. As validation for our approach, we compare our detected inflation event at Akutan with prior work by \citet{Ji2011}.

Complications in continuous GPS position measurements necessitate first preprocessing and conditioning of the data. In particular, the extensive spatial scale of the PBO network makes manual inspection of volcanoes overly time consuming, while the GPS instruments themselves suffer from spatially \citep{Dong2006} and temporally \citep{Langbein2008} correlated noise that could mask signals of interest. The computer aided discovery approach helps overcome these inherent challenges in analyzing the GPS time series data by providing a framework for uniformly preprocessing to reduce temporally correlated noise and remove secular drifts and seasonal trends. Our framework also supports independent and parallelizable runs examining multiple points of interest to first exclude large numbers of sites without events and then present a more tractable set of points of interest for further  examination. These parallel runs can be distributed locally as multiple processes on a multi-core machine or offloaded to servers in the Amazon Web Services cloud \citep{AWS2015}.

\begin{figure}[h]
  \centering
    \includegraphics[width=\textwidth]{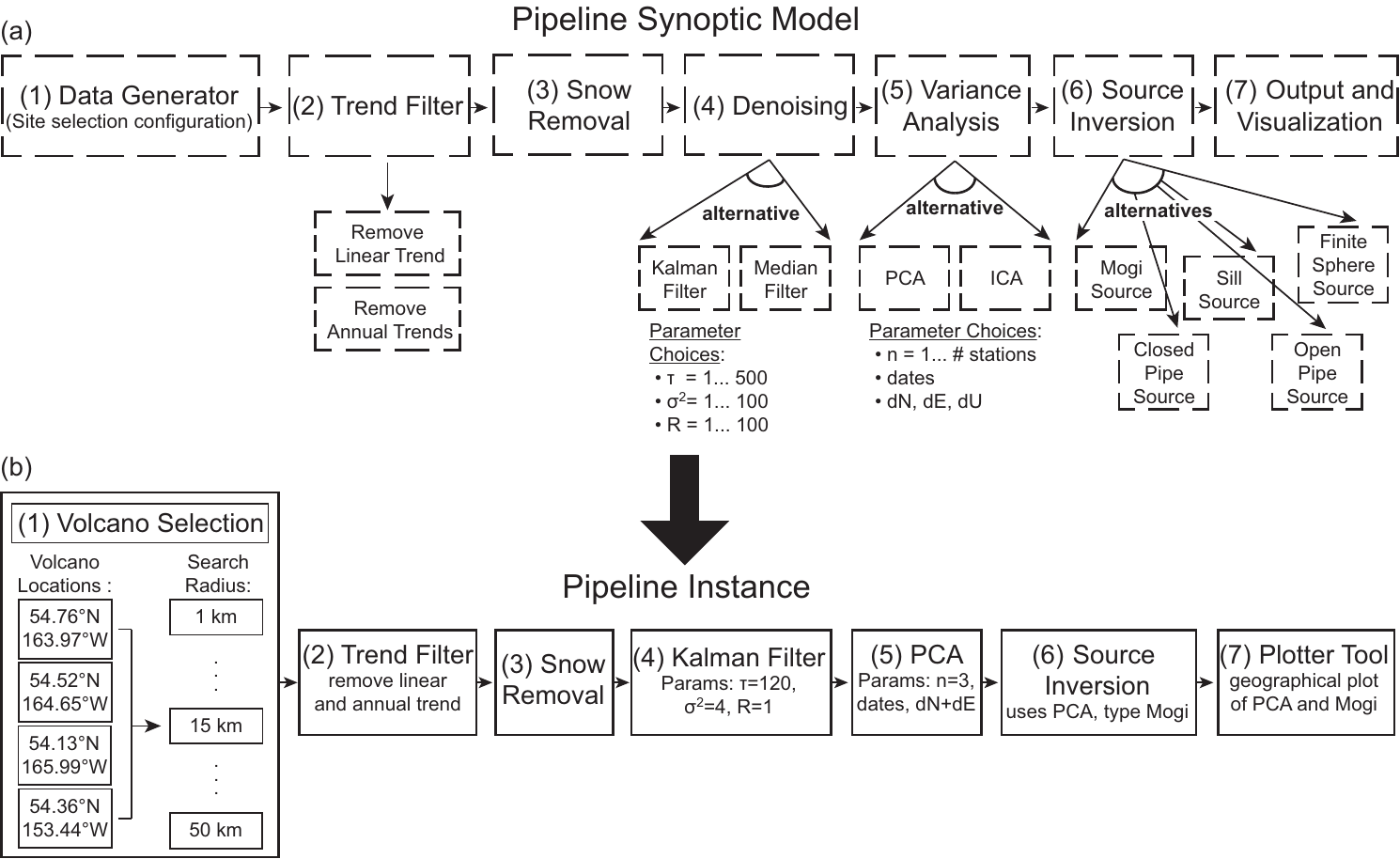}
    \caption{(a) A configurable pipeline meta-model with scientist-defined algorithmic choices and parameter range choices for the geodetic time series data in this study. (b) A specific computer-assisted pipeline instance that runs over a range of geographic windows of interest for detecting and modeling transient volcanic events.}
    \label{fig:Fig2}
\end{figure}

The computer aided discovery system utilized here is an implementation under continuing development \citep{Pankratius2016}. The overall approach is to create configurable data processing frameworks that then generate a specific analysis pipeline configuration. Figure~\ref{fig:Fig2}(a) exemplifies our pipeline synoptic model, which is a meta-model summarizing possible pipelines that make scientific sense for our particular GPS data and analysis goals. Choices for processing stages and the parameter ranges for each stage can then be selected as desired for a particular pipeline instance. For example, the notation at stage ``(4) Denoising'' means that the general denoising step in the pipeline can be implemented by either choosing a Kalman filter or a median filter. The notation for an ``alternative'' choice is based on the notation commonly used in the generative programming community \citep{Czarnecki2000}. If a Kalman filter is chosen, then its three parameters can be chosen from the following intervals: $\tau\in[1,500]$,  $\sigma^2\in[1,100]$, and $R\in[1,100]$, where $\tau$ is a correlation time in days, $\sigma^2$ is the standard deviation of the correlated noise, and $R$ is the measurement error. The pipeline is structured for modular flexibility, where the Kalman smoother stage can be replaced with any other smoothing filter such as a median or low pass filter. Similarly, while Principal Component Analysis (PCA) is a commonly used analysis technique for geodetic analysis and inversion \citep{Dong2006, Kositsky2010, Lin2010}, this framework allows for an easy substitution with other approaches such as an Independent Component analysis (ICA) \citep{Hyvarinen2000}.

Figure~\ref{fig:Fig2}(b) shows the particular pipeline derived from the general model in Figure~\ref{fig:Fig2}(a) which we then use throughout our analysis. In our Python implementation, this architecture is realized by means of interlinked ``pipeline stage containers'' that hold the actual processing components as Python objects (e.g. filters). For this pipeline, the system varies the radius around a geographic region of interest to include additional or fewer GPS stations for sensitivity analysis. The stage containers operate on the data in place and thus reduce read and write complexities and storage requirements, to allow for the analysis of large amounts of data.

The first stage of our implemented pipeline (Figure~\ref{fig:Fig2}(b)) queries the combined snow and GPS data from a data generator object and represents the data source for the pipeline. The data generator takes as input parameters the AVO list of volcanoes and selects PBO GPS stations in their vicinity. It also applies a procrustean method for estimating up to 3 translation, 3 rotation, and 1 scale parameters to compute the transformation to the local Alaskan reference frame using all of the PBO sites in Alaska to reduce the common mode error \citep{Awange2008}. For volcanoes with a sufficient number of GPS stations (at least 3 stations), the pipeline then applies a detrending filter (Stage 2) that removes linear drift and annual and semi-annual sinusoidal trends, a stage (Stage 3) that removes stations with excessive errors due to snow, a denoising stage (Stage 4) that applies a Kalman smoother (as implemented via the Fraser-Potter approach by \citet{Ji2011} and whose details can be found in \citet{Ji201D}[Appendix A,B]) to reduce the data noise and improve the signal-to-noise ratio, and a principal component analysis (PCA, Stage 5) to identify the overall direction and magnitude of motion of the GPS stations. Finally, the pipeline fits a physics-based source model (Stage 6) and provides a visualization (Stage 7) of the PCA components and the fit from the selected model.

This modular pipeline approach allows for the easy substitution and comparison of different types of filters and analysis techniques over a range of parameter values. A given pipeline instance can also be easily reused for a variety of application scenarios. Preconfigured processing pipelines with parameters selected for positive detections on certain datasets can be applied to validate detections for other volcanoes and to make new discoveries using newly acquired data sets. The standardized assembly and parameterization of this pipeline approach will also be useful for ensuring reproducibility of detections.

\section{A Scalable Approach to Detecting Volcanic Transient Deformation Events}

\begin{figure}[h]
  \centering
    \includegraphics[width=\textwidth]{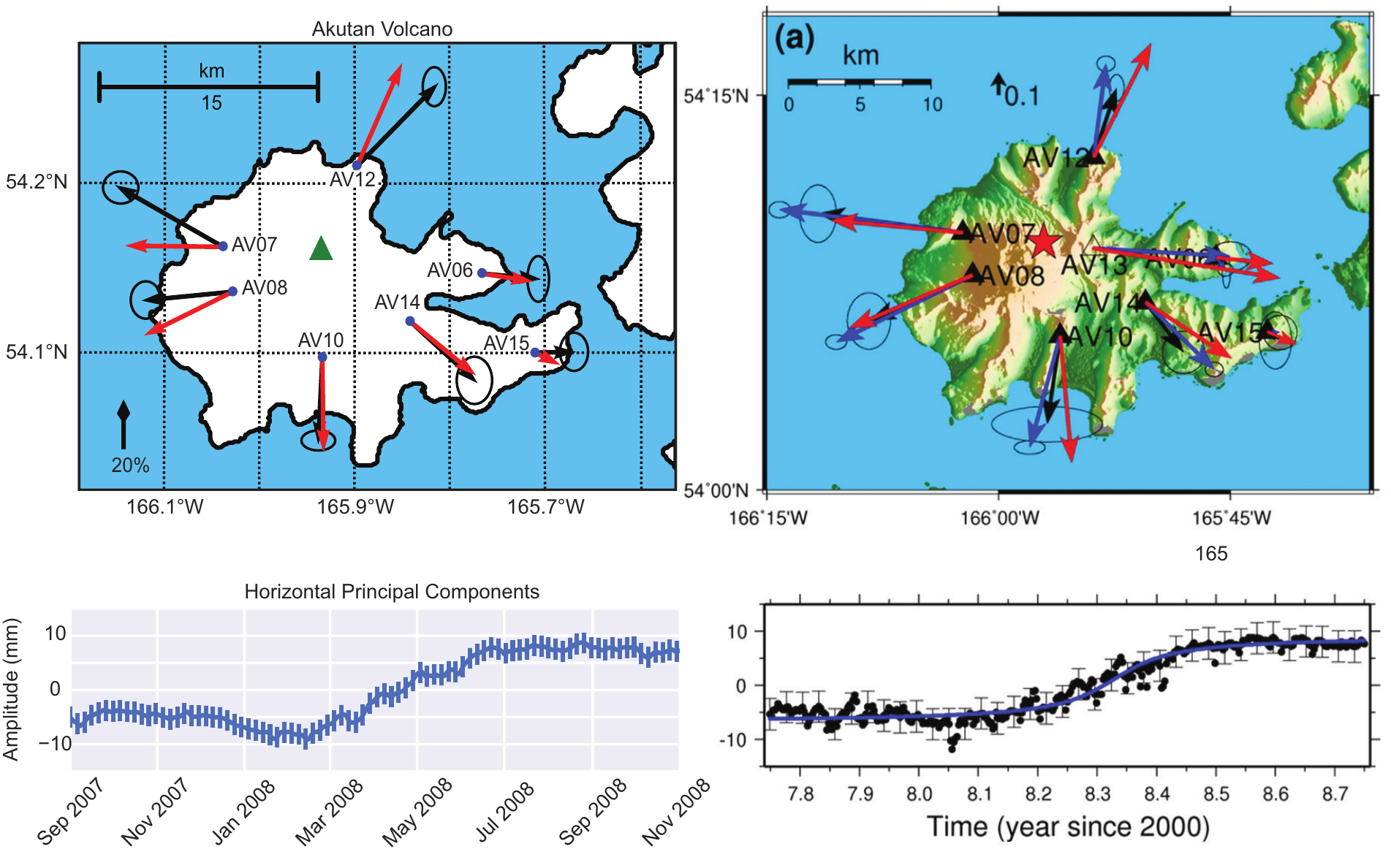}
    \caption{Transient inflation event validation. Left column: detection of the transient event at Akutan Island, Alaska in 2008 by the computer aided discovery pipeline in this study; right column: detection as presented by \citet{Ji2011}. The top row shows the spatial motions of the first principal component (black) along with error ellipses and the spatial Mogi model fit (red). The 20\% and 10\% arrows in the plots indicate the scaling of the vectors relative to the first horizontal PCA component's amplitude. The bottom row shows the temporal pattern of the first principal component, with 1-sigma error bars in blue on left and in black on right.}
    \label{fig:Fig3}
\end{figure}

In this exploratory study, we apply our technique over all 137 volcanoes listed by the AVO for the time interval between 1/1/2005 and 1/1/2015. We first run a preliminary version of our pipeline and generate a PCA eigenvector geographic plot for each volcano with at least 3 GPS stations within a 20 km radius. While volcanoes with insufficient number of GPS stations are automatically discarded, we visually inspect the eigenvector plots to exclude volcanoes with incorrectly clustered stations. The horizontal motion PCA projections are then plotted for the stations at the remaining volcanoes, from which we identify possible inflation events based on the observed horizontal motion and specify a narrower time window around these potential transient events. Finally, we run our complete analysis pipeline on these volcanoes with the PCA analysis stage (Stage 5) taking as parameters the more specific dates and examine the resulting horizontal principal component (PC) projection to identify the transient inflation event (e.g. left panels of Figure~\ref{fig:Fig3}). We also fit a Mogi source model (Stage 6) to estimate the likely location of the magma chamber and compare the model deformation with the principal component motion (e.g. matching the red and blue displacement vectors in the upper left panel of Figure~\ref{fig:Fig3}).

Of the 137 volcanoes analyzed in the 2005 to 2015 time span, we find 9 volcanoes with at least 3 GPS stations nearby, although only 4 volcanoes have sufficient data and GPS stations for further analysis after removing stations with significant snow errors. While this step of selecting volcanoes with sufficient GPS stations and data is nominally trivial, we include this step to describe our complete analysis pipeline which will aid in the reduction of the potential search space resulting from increasingly large and dense ongoing and future datasets. From examining these 4 volcanoes, Akutan (Akutan Island), Augustine (Augustine Island), Shishaldin (Unimak Island), and Westdahl (Unimak Island), we discover 3 transient inflation events: in early 2008 at Akutan, in early 2011 at Westdahl, and in early 2013 at Shishaldin. In particular, the results from the 2008 inflation event at Akutan are useful for validating our computer aided pipeline approach, as this inflation event was previously detected by \citet{Ji2011}. Figure~\ref{fig:Fig3} compares our results with the work by \citet{Ji2011} and shows a strong match in the direction of the eigenvectors and in the overall amount of detected inflation. The differences in the projected principal component motion and the Mogi source motion are not significant and can be attributed to differences in the specific implementation of the reference frame stabilization method, the Kalman smoothing parameters, and the mechanics of our method for fitting the magma chamber location.

This analysis pipeline can be easily reused to examine GPS motion associated with other volcanoes monitored by AVO. We describe here results from analysis of Augustine, Westdahl, and Shishaldin. During this 10 year time span, between 2005 and 2015, no transient events were observed at Augustine, matching the corresponding recorded activity log obtained from AVO. Although AVO noted an eruption at Augustine in 2006, the available stations used in this analysis were not installed until after the eruption, as a replacement for stations destroyed by that event.

\begin{figure}[!]
  \centering
    \includegraphics[width=\textwidth]{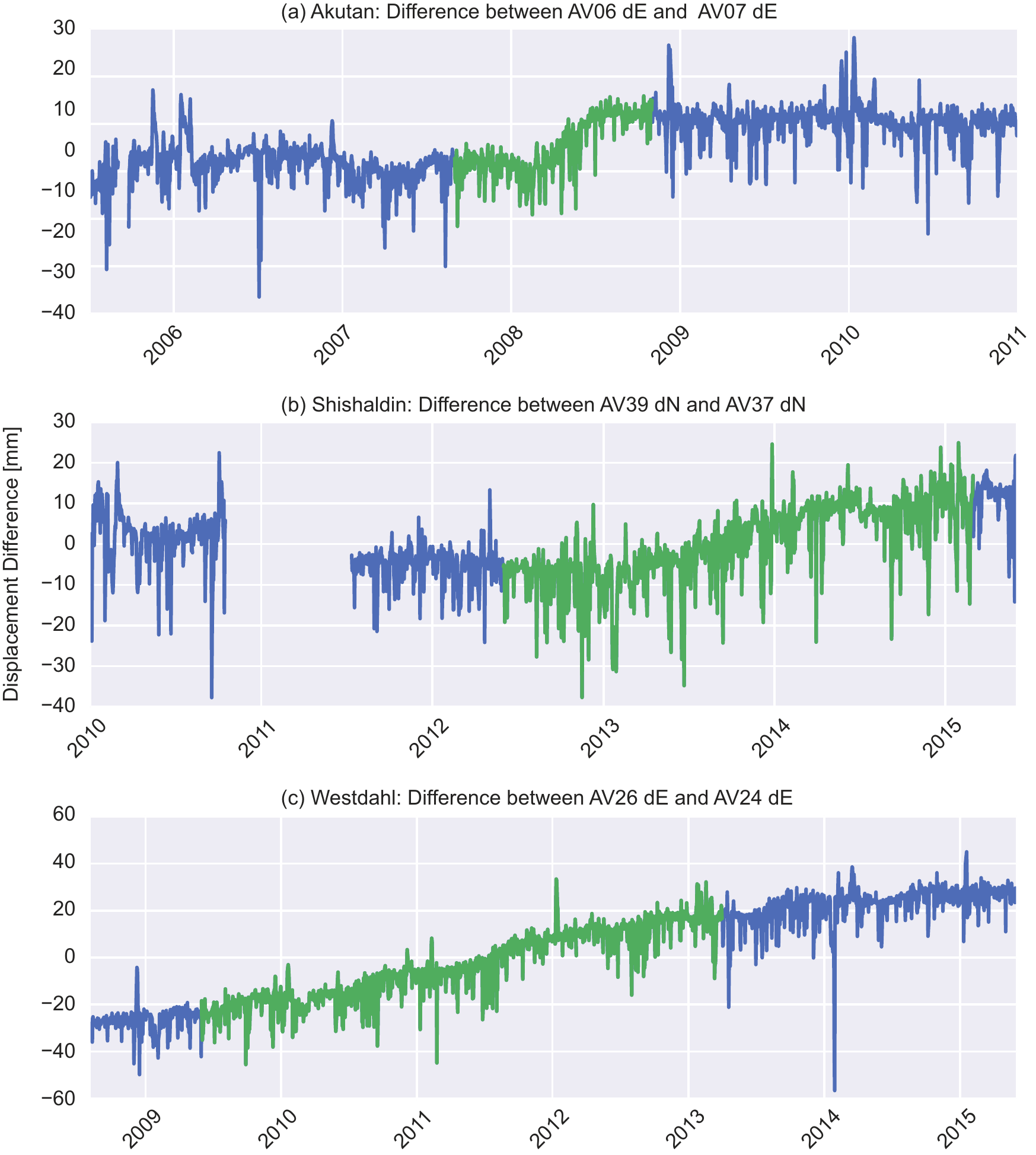}
    \caption{Validation of transient inflation events by comparison of the raw difference between (a) the east-west (dE) motion at stations AV06 and AV07 at Akutan, (b) the north-south (dN) motion at AV39 and AV37 at Shishaldin, and (c) the east-west (dE) motion at stations AV26 and AV24 at Westdahl. The stations are located on opposite sides of the volcanoes and, as they move further apart, show the volcano is inflating. The green segments mark the portions of data processed and described by the PCA. The full PCA results for Akutan are shown in Figure~\ref{fig:Fig3}, for Shishaldin in Figure~\ref{fig:Fig5}(a), and for Westdahl in Figure~\ref{fig:Fig5}(b).}
    \label{fig:Fig4}
\end{figure}

We detected two transient inflation events that have not been previously reported in the literature, one each at Westdahl and at Shishaldin, by noting the sudden occurrence of inflation in the first principal component. We also verify that these are indeed transient events by comparing the unprocessed motions from stations on opposite sides of the volcano moving in opposite directions (N-S or E-W), as shown in Figure~\ref{fig:Fig4}. This verification identifies the signal seen at Akutan, Figure~\ref{fig:Fig4}(a), by a comparison of the difference in E-W motion between AV06 and AV07 (located on opposite outer ends of the volcano), and also validates the two new transient events observed at Westdahl and Shishaldin, Figures~\ref{fig:Fig4}(b) and (c).

\begin{figure}[h!]
  \centering
    \includegraphics[width=\textwidth]{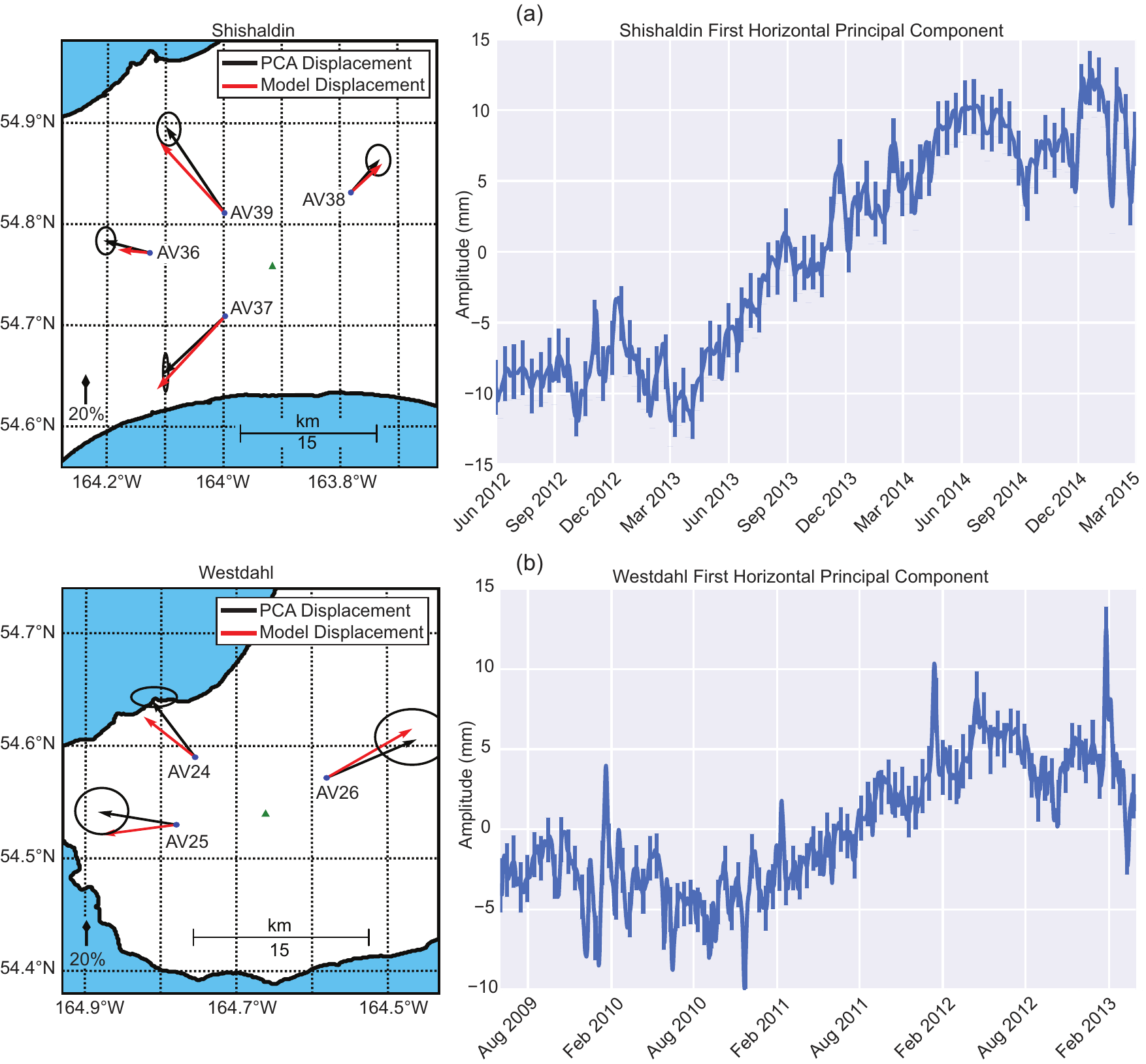}
    \caption{Observations of newly detected transient inflation events (a) from early 2013 and to early 2015 at Shishaldin volcano on Unimak Island, Alaska, and (b) from early 2011 to 2012 at Westdahl volcano on Unimak Island, Alaska. Note that some of the spikes in the data are likely artifacts due to snow. Two additional GPS stations were available for Westdahl in this time region, but were not used due to large portions of the data being corrupted by snow. The 20\% arrows in the plots indicate the scaling of the vectors relative to the respective first horizontal PCA component's amplitude.}
    \label{fig:Fig5}
\end{figure}

For Shishaldin, we observe a transient inflation event beginning in early 2013 and lasting until early 2015, with about 20 mm of horizontal displacement as shown in Figure~\ref{fig:Fig5}(a). The first horizontal PC explains ~65\% of the variance in the data. For Westdahl, we observe a transient inflation event beginning in early 2011 and ending in 2012, with about 8 mm of horizontal displacement as shown in Figure~\ref{fig:Fig5}(b). The first horizontal PC explains ~63\% of the variance in the data.

\section{Discussion}
\subsection{Transient Inflation Events}

Of the 4 volcanoes with sufficient GPS stations to conduct analysis, we detect one previously reported inflation event at Akutan, two new events at Shishaldin and Westdahl, and no recent events at Augustine. According to the AVO logs of volcanic activity for these four volcanoes, no other volcanic events occurred at these volcanoes during the 2005 to 2015 time period for our available data. While transient deflation events are also important to study, in characterizing other behavior of magma activity, and could be detected with our analysis, all of the events identified in this study are inflation events.

The two new inflation events at Shishaldin and Westdahl occurred at Unimak Island, which is the easternmost Aleutian Island and the home to a number of volcanoes with significant amounts of activity. In the last two and a half centuries, Shishaldin has had 63 reported events, and Westdahl 13 events \citep{AVO_web}. The island has been under considerable study due to its multiple volcanoes, including Shishaldin, Westdahl, Pogromni, Fisher, Isanotski, and Roundtop, and its persistent seismic and volcanic activity \citep{Mann2003, Lu2003, Gong2015}. The two transient events identified in this study using GPS observations occur simultaneously with AVO reports of volcanic activity.

The 2011 inflation at Westdahl may correspond with AVO observations in mid-2010 of seismic activity at Westdahl, although no eruption was reported. The event persists throughout 2012 with potentially varying changes in the rate of inflation. This behavior is consistent with Interferometric Synthetic Aperture Radar (InSAR) and GPS observations and with modeling by \citet{Gong2015} showing that Westdahl has been inflating at changing rates over at least the past decade. Similarly, the 2013 inflation we detect at Shishaldin matches with AVO observations in February 2013 of seismic events. And, as the inflation event persisted until early 2015, the inflation event may also be connected with an effusive eruption in 2014 and with later reports of seismic and steam activity at Shishaldin that persisted through 2015. While the analysis of stations located on opposite sides of the volcanoes in Figure~\ref{fig:Fig4} shows a clear transient signal observed at Akutan and Shishaldin, the signal at Westdahl is less clear. This is likely due to the limitations of stations AV26 and AV24 at Westdahl not being located on directly opposite sides of the volcano, resulting in a noisier difference.

\subsection{Advantages of Configurable Computer-Aided Volcanic Data Processing Pipelines}

To our knowledge, neither of these inflation events have been previously described in other works, demonstrating the capability of our newly described computer aided discovery system for assisted rapid discovery. This approach differs from previous work such as \citet{Ji2011}, which considered the entire region of Alaska and observed the strong eigenvector motion only at Akutan. Rather than attempting to detect an exceptionally strong signal out of the background noise, which risks missing smaller events, this pipeline approach targets known locations of interest and analyzes these regions at varying spatial scales. This multiscale approach can be successively applied to larger geographic areas by specifying grids of interest and running the pipeline on time series obtained from the respective GPS stations. Potentially, this approach may also enable real-time or near real-time detection of future inflation events by continuous analysis of real-time GPS observations.

\begin{figure}[!ht]
  \centering
    \includegraphics[width=\textwidth]{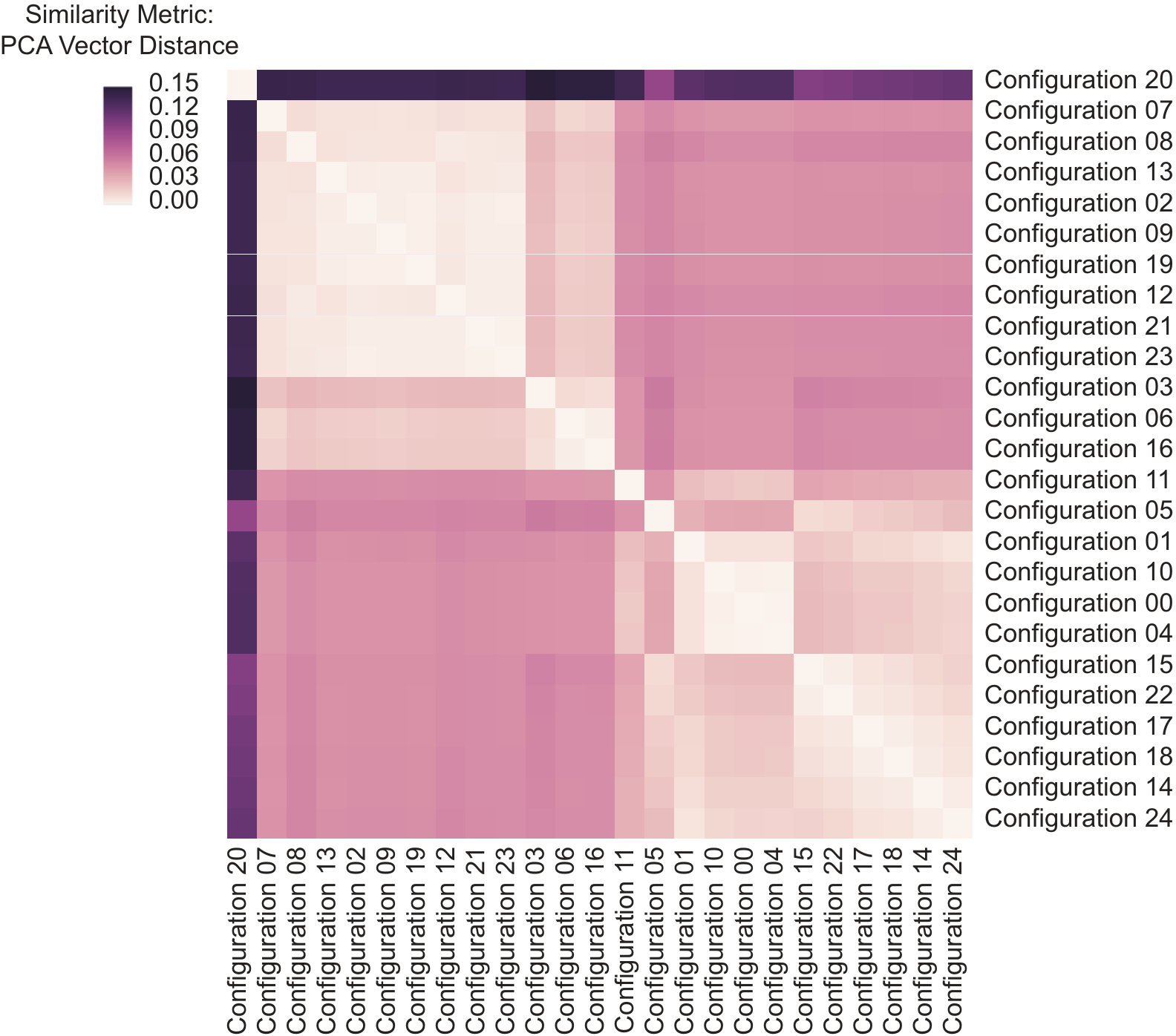}
    \caption{A heatmap showing the sum of the euclidean distance between the eigenvectors at each PBO station for each sensitivity analysis run at Akutan. Colors reflect the distance between results from different configurations as measured by the total difference between PCA eigenvectors at each PBO station. Visualization of configurations provides natural groupings of potential model assumptions for the given Akutan empirical data set. The actual parameters for each configuration are provided in the \hyperref[sec:Appendix]{Appendix}}
    \label{fig:Fig6}
\end{figure}

A particular output provided by our framework is a heatmap (Figure~\ref{fig:Fig6}) that allows comparison of results from different pipeline configurations, where each configuration makes different assumptions about underlying physics. For our Akutan case, we demonstrate here an example of this functionality by testing 25 different configurations with implicitly different noise properties as addressed by Kalman smoothers with certain parameters or a median filter with varying window lengths. The results in Figure~\ref{fig:Fig6} are compared using a color scheme that represents the difference between configurations as measured by the sum of all Euclidean distances between the eigenvectors obtained at each PBO station.

The colors in the heatmap in essence describe ``how different'' pipeline results are given various assumptions. The colors indicate groups of similar configurations, which could help structure ideal configurations for further studies. This output can reveal visual hints for potential alternative models that might fit the same empirical observations with different algorithmic choices and parameters. In our particular numerical example for Akutan, however, the differences even between the maximum and minimum parameter values were minimal, so we simply present the results from the initial configuration, Configuration 00. (The full list of configuration parameters can be found in the \hyperref[sec:Appendix]{Appendix}.) This capability to vary configuration parameters extends to all stages of the pipeline, including adjusting the details of the reference frame stabilization to remove common mode errors in the data generator stage (Stage 1, Figure~\ref{fig:Fig2}).

\subsection{Fitting Magma Chamber Model Variants}

\begin{figure}[!ht]
  \centering
    \includegraphics[width=\textwidth]{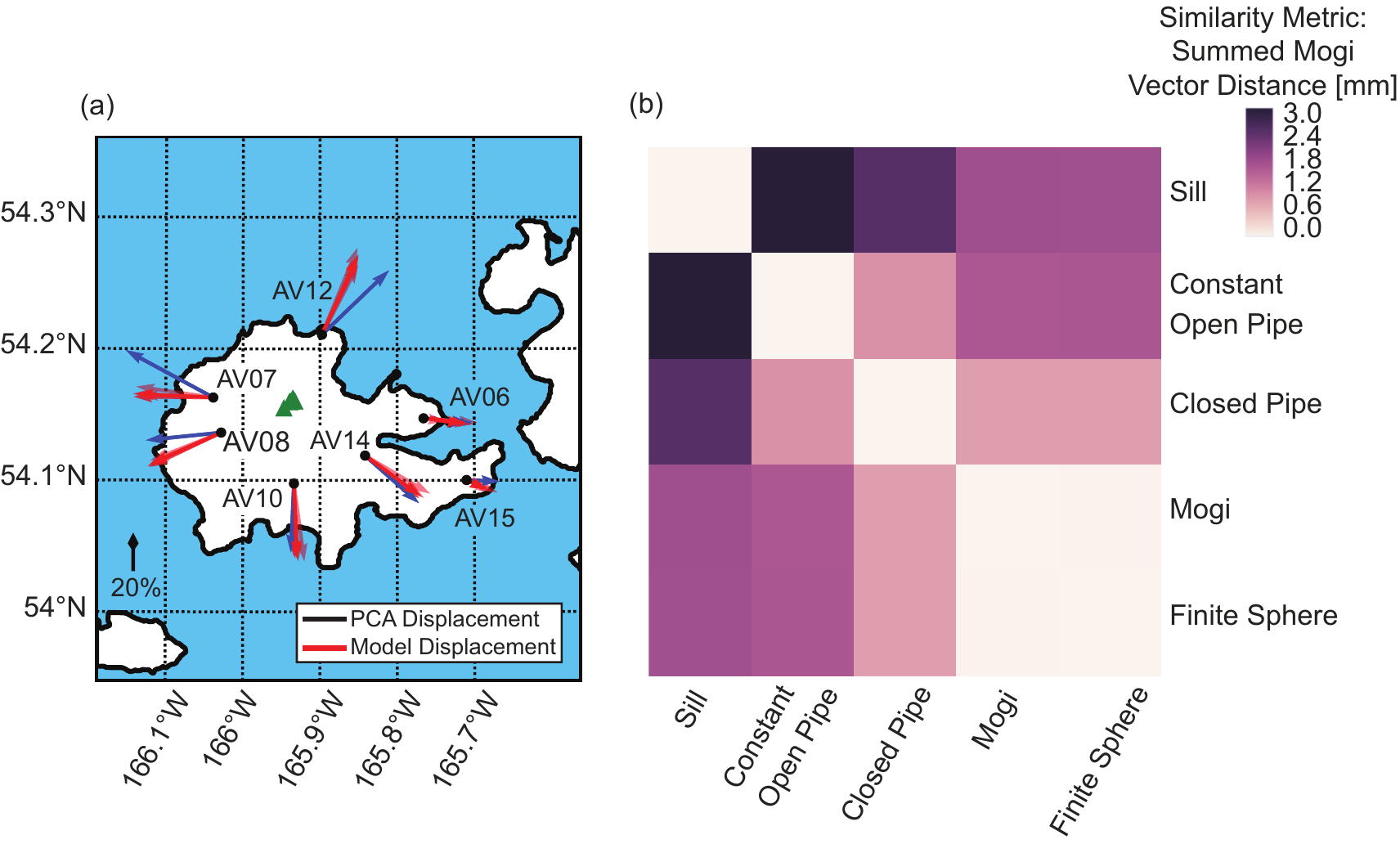}
    \caption{Comparison of 5 different source model inversions -- Mogi, finite sphere, closed pipe, constant (width) open pipe, and sill -- for the motion described by the eigenvectors for Akutan displayed on (a) a geographical plot with the eigenvectors in blue, the different inversion displacements in red (the solid color is the mean), and the different magma source locations in green (the solid color shows the average location). The 20\% arrow indicates the scaling of the vectors relative to the respective first horizontal PCA component's amplitude.(b) A heatmap that compares the summed Euclidean distance [mm] between model displacement vectors at each GPS station for different model sources}
    \label{fig:Fig7}
\end{figure}

We combine the output from the PCA analysis with the smoothed detrended GPS position time series to determine station displacements, which can be used in the model inversion analysis stage to fit a magma chamber source associated with the detected inflation event. Using another set of configurations, we can test different model variants to evaluate a best fit that describes the observed 3-dimensional surface displacements as characterized with arctangent parameterizations. Here, we fit the standard Mogi point source model \citep{Mogi1958} and then evaluate the effect of modeling the magma chamber at Akutan as a finite sphere, a closed pipe, a constant (width) open pipe, or a sill-like magma source \citep{Dzurisin2006volcano}. The results of testing different magma source types is shown in Figure~\ref{fig:Fig7}. The geographic plot in Figure~\ref{fig:Fig7}(a) reveals that the source location is mostly stable across the different models. As measured by the heatmap in Figure~\ref{fig:Fig7}(b), only the sill-like magma source results in a less similar model, and even then the difference is not substantial with at most a combined difference of around 3 mm between the projected displacement vectors at all PBO stations as compared with around 30 mm of total motion. This minimal difference is also reflected by the mostly-negligible difference in projected displacement vectors shown in red in Figure~\ref{fig:Fig7}(a). Further work could account for additional model types, as well as more complex combinations of magma sources.

\subsection{Instrument Sensor Network Requirements and Future Trends}

The four Alaskan volcanoes examined in this study, Akutan, Augustine, Westdahl, and Shishaldin, have a sufficiently dense network of GPS receivers for running PCA, but Westdahl demonstrates a greater detection uncertainty due to its less dense network of GPS stations. The verification of the transient event in Figure~\ref{fig:Fig4}(c) is less distinct, and the error ellipses in Figure~\ref{fig:Fig5}(b) are larger than those in Figure~\ref{fig:Fig3} for Akutan and Figure~\ref{fig:Fig5}(a) for Shishaldin. This higher uncertainty in the case of Westdahl, or the inability to conduct further analysis as in the case of other Alaskan volcanoes listed by the AVO is often due to the reduced number of available GPS stations after removing stations with corrupted data. This data loss,  which results from the accumulation of snow and ice on the GPS antenna, could be mitigated by using an improved filter that better corrects for the corrupted data or by more accurately removing only the corrupted sections of data, instead of the current approach of rejecting the station entirely. As an example, in the case of Westdahl, two additional GPS stations, which were excluded from this study when examining the combined PBO and NSIDC data, might become useable.

Consequently, our approach is limited by the amount of available data at any given location, requiring volcanoes, or other points of interest, to be instrumented with a sufficiently dense network of sensors. However, this work looks forward towards the decreasing cost of sensors and the trend of increasing density of GPS stations and the resulting increase in volume of data. As larger and higher resolution sensor networks are deployed across the entire planet, traditionally manual approaches become inefficient, making Big Data analysis a key bottleneck in the scientific discovery process. This work demonstrates that our approach can enable scientists to study more difficult problems by better utilizing large amounts of data.

A key capability of an automated assisted system is the ability to aid in downloading, parsing, and analyzing increasingly large data sets. Given a broad range of scientific applications using GPS data, we downloaded the data set for all PBO GPS stations ($\sim$1GB). From that larger data set, when we focus on exploring events associated with volcanic activity, our pipeline automatically narrows down the relevant GPS stations based on the provided coordinates of interest. Furthermore, as different types of instruments cover the same geographical region of interest, this pipeline framework also helps integrate additional sources of data. Here, the pipeline preprocesses and integrates the snow cover NSIDC data (an additional 150 GB data set). And as more GPS stations are established and additional data sources such as InSAR are integrated into a cohesive analysis, the computational footprint for data analysis will continue to increase.

The flexibility of our framework allows it to be adapted to data sets as they become available. Although the transient inflation events discussed here have been on the order of months to years, with higher time resolution data of the same precision, shorter events could also be detected. Furthermore, while the motion that characterizes transient events only becomes detectable after a period of time, real-time or near real-time monitoring can be implemented to aid in providing early warning of volcanic activity.

\section{Conclusion}

This article introduces a computer-aided discovery approach that develops a framework for studying transient events at volcanoes using configurable data processing pipelines. We demonstrate its applicability and utility by detecting and analyzing transient inflation events at 3 Alaskan volcanoes, Akutan on Akutan Island and Westdahl, and Shishaldin on Unimak Island. Our results include the discovery of two inflation events that have not been previously reported in the literature.

The study of transient events at Alaskan volcanoes exemplifies how volcanic data exploration can be a considerable challenge for scientists due to the large search space of potential events and event features hidden within large amounts of data. Computer assistance and automation expedites the analysis of multi-scale transient deformation events at a large number of sites of interest; our system is assembled from easily reusable modular analysis stages, providing an extensible structure that can be used to explore multiple deformation hypotheses efficiently. In the future, we envision that such techniques will become a key component in processing the data collected from ever growing sensor networks that produce increasingly large amounts of data. In addition, these computational pipelines can be expanded to take advantage of data fusion from other data sources such as InSAR, to aid in the detection and analysis of earth surface deformation and volcanic events.

While our overall computer-aided discovery framework is still under continued revision and development, we have released our Python package for accessing various scientific datasets online as the SciKit Data Access (skdaccess) package. This package can be easily installed through the Python Package Index (https://pypi.python.org/pypi/skdaccess/), and the source code is available on GitHub (https://github.com/skdaccess/skdaccess). We plan to release the analysis package incrementally in a similar fashion. To promote the importance of open code and open science, usage of our framework and code will be covered under the MIT Open License.

\section*{Acknowledgements}

We acknowledge support from NASA AIST NNX15AG84G, NSF ACI 1442997, and the Amazon Web Services Research Grant. This material is based on EarthScope Plate Boundary Observatory data services provided by UNAVCO through the GAGE Facility with support from the National Science Foundation (NSF) and National Aeronautics and Space Administration (NASA) under NSF Cooperative Agreement No. EAR-1261833.

\section*{Appendix}
\label{sec:Appendix}

The heatmap in Figure 6 is generated from multiple executions of the pipeline using the configurations listed below, where the filtering/smoothing stage was perturbed. For the Kalman filter stage container, the three parameters are $\tau$ (correlation time in days), $\sigma^2$ (variance of the correlated noise), and $R$ (measurement noise). For the median filter, the one parameter is the window length (the number of days of the window for calculating the median). To give two examples from our 25 configurations, Configuration 0 employs a Kalman filter with a $\tau$ of $120$, a $\sigma^2$ of $4$, and an $R$ of $1$, and Configuration 2 employs a median filter with a window length of $79$ days.
\\

Configuration 0: KalmanFilter1['120', '4', '1']

Configuration 1: KalmanFilter1['65', '99', '65']

Configuration 2: MedianFilter['79']

Configuration 3: MedianFilter['23']

Configuration 4: KalmanFilter1['191', '46', '8']

Configuration 5: KalmanFilter1['417', '19', '86']

Configuration 6: MedianFilter['34']

Configuration 7: MedianFilter['51']

Configuration 8: MedianFilter['98']

Configuration 9: MedianFilter['70']

Configuration 10: KalmanFilter1['279', '76', '11']

Configuration 11: KalmanFilter1['500', '100', '1']

Configuration 12: MedianFilter['88']

Configuration 13: MedianFilter['69']

Configuration 14: KalmanFilter1['241', '86', '89']

Configuration 15: KalmanFilter1['500', '51', '96']

Configuration 16: MedianFilter['32']

Configuration 17: KalmanFilter1['421', '48', '56']

Configuration 18: KalmanFilter1['60', '5', '30']

Configuration 19: MedianFilter['75']

Configuration 20: KalmanFilter1['500', '1', '100']

Configuration 21: MedianFilter['87']

Configuration 22: KalmanFilter1['207', '26', '95']

Configuration 23: MedianFilter['83']

Configuration 24: KalmanFilter1['225', '52', '34']

\end{document}